\documentstyle[prl,twocolumn,aps,english]{revtex}

\def\beq{\begin{equation}}
\def\eeq{\end{equation}}
\def\6{\langle}
\def\9{\rangle}
\def\half{\mbox{$1\over2$}}
\def\bn{\mbox{\boldmath$n$}}
\def\nJ{\mbox{\boldmath$n\cdot J$}}
\def\dtp{d_{\theta\phi}}

\begin{document}
\draft

\title{Entangled Quantum States as Direction Indicators}

\author{Asher Peres and Petra F. Scudo}
\address{Department of Physics, Technion---Israel Institute of
Technology, 32000 Haifa, Israel}

\maketitle
\begin{abstract}
We consider the use of $N$ spin-\half\ particles for indicating a
direction in space. If $N>2$, their optimal state is entangled. For large
$N$, the mean square error decreases as $N^{-2}$ (rather than $N^{-1}$
for parallel spins).
\end{abstract}

\bigskip
\pacs{PACS numbers: 03.67.Hk, 03.65.Ta}

Information theory usually deals with the transmission of a sequence
of discrete symbols, such as 0 and 1. Even if the information to
be transmitted is of continuous nature, such as the position of a
particle, it can be represented with arbitrary accuracy by a string
of bits. However, there are situations where information cannot be
encoded in such a way. For example, the emitter (conventionally called
Alice) wants to indicate to the receiver (Bob) a direction in space.
If they have a common coordinate system to which they can refer, or
if they can create one by observing distant fixed stars, Alice simply
communicates to Bob the components of a unit vector \bn\ along
that direction, or its spherical coordinates $\theta$ and $\phi$. But
if no common coordinate system has been established, all she can do is
to send a real physical object, such as a gyroscope, whose orientation
is deemed stable.

In the quantum world, the role of the gyroscope is played by a system
with large spin. For example, Alice can send angular momentum eigenstates
satisfying $\nJ|\psi\9=j|\psi\9$. This is essentially the solution
proposed by Massar and Popescu~\cite{mp} who took $N$ parallel spins,
polarized along~\bn. The fidelity of the transmission is usually defined
as

\beq F=\6\cos^2(\chi/2)\9=(1+\6\cos\chi\9)/2, \eeq
where $\chi$ is the angle between the true \bn\ and the direction
indicated by Bob's measurement. The physical meaning of $F$ is that
$1-F=\6\sin^2(\chi/2)\9$ is the mean square error of the measurement,
if the error is defined as $\sin(\chi/2)$. The experimenter's aim,
minimizing the mean square error, is the same as maximizing fidelity. We
can of course define ``error'' in a different way, and then fidelity
becomes a different function of~$\chi$ and optimization leads to
different results~\cite{m}. Here, we shall take Eq.~(1) as the
definition of fidelity.

Massar and Popescu showed that for parallel spins, $1-F=1/(N+2)$. It then
came as a surprise that for $N=2$, parallel spins were not the optimal
signal, and a slightly higher fidelity resulted from the use of opposite
spins~\cite{gp}. The intuitive reason given for this result was the use
of a larger Hilbert space (four dimensions instead of three). This raises
the question what is the most efficient signal state for $N$ spins, whose
Hilbert space has $2^N$ dimensions. Will $F$ approach~1 exponentially? In
this Letter, we show that the optimal result is a quadratic approach,
as illustrated in Fig.~1.  

Our first task is to devise Bob's measuring method, whose mathematical
representation is a positive operator-valued measure (POVM)~\cite{qt}.
For any unit vector \bn, not necessarily Alice's direction, let
$|j,m(\bn)\9\equiv|j,m(\theta,\phi)\9$ denote the coherent angular
momentum state~\cite{perel} that satisfies

\beq \mbox{\boldmath$J$}^2\,|j,m(\bn)\9=j(j+1)\,|j,m(\bn)\9, \eeq
and
\beq \nJ\,|j,m(\bn)\9= m\,|j,m(\bn)\9. \eeq
We then have~\cite{perel}

\beq (2j+1)\int\dtp\,|j,m(\theta,\phi)\9\6j,m(\theta,\phi)|
  ={\bf1}_j, \label{1j}\eeq
where 

\beq \dtp:=\sin\theta d\theta\,d\phi/4\pi, \eeq
and ${\bf1}_j$ is the projection operator over the $(2j+1)$-dimensional
subspace spanned by the vectors $|j,m(\theta,\phi)\9$. If $N=2$, so that
$j$ is 0 or 1, the two resulting subspaces span the whole 4-dimensional
Hilbert space. For higher $N$, all the rotation group representations
with $j<N/2$ occur more than once.  We then have, if we take each $j$
only once, from 0 or \half\ to $N/2$,

\beq \sum(2j+1)={(N+2)^2\over4}\quad\mbox{or}
  \quad {(N+1)(N+3)\over4},\eeq
for even or odd $N$, respectively. For large $N$, the dimensionality of
the accessible Hilbert space tends to $N^2/4$, and this appears to be
the reason that the optimal result for $1-F$ is quadratic in $N$, not
exponential. An intuitive argument for this quadratic behavior was given
by Aharonov and Popescu~\cite{ap}. No improvement results if we endow
the particles with internal quantum numbers such as charge or
strangeness, so that the entire Hilbert space can be spanned by states
with distinguishable properties, because any additional information
that Alice could send to Bob would refer to these new quantum numbers,
not to the direction of \bn.

We now turn to the construction of Bob's POVM~\cite{qt}.  Let $\rho$
denote the initial state of the physical system that is measured. All
these input states span a subspace of Hilbert space.  Let {\bf1} denote
the projection operator on that subspace. A POVM is a set of positive
operators $E_\mu$ which sum up to {\bf1}. The index $\mu$ is just a
label for the outcome of the measuring process. The probability of
outcome $\mu$ is $\mbox{tr}(\rho E_\mu)$.  In the present case, $\mu$
stands for the pair of angles $\theta\phi$ that are indicated by Bob's
measurement. If we want a high accuracy, these output angles should have
many different values, spread over the unit sphere~\cite{polyhed}. For
example, the components of a continuous POVM, as in Eq.~(\ref{1j}), are
given by

\beq E_{\theta\phi}=(2j+1)\,\dtp\,|j,m(\theta,\phi)\9\6j,m(\theta,\phi)|.
 \eeq
Such a POVM with $m=j$ corresponds to the method of Ref.~\cite{mp}. The
choice $m=j$ is not optimal. As shown in~\cite{gp} for the case $N=2$,
signal states with opposite spins give a higher fidelity. With our
present notations, these states are $(|0,0\9+|1,0(\bn)\9)/\sqrt2$. They
involve two values of $j$, but a single value of $m$, namely~0.

One possibility to include several values of $j$ in a POVM is to take a
sum of expressions like (\ref{1j}). This brings no advantage, because a
convex combination of POVMs cannot yield more information than the best
one of them~\cite{cnvx}. Optimal POVM components can always be assumed
to have rank one. Therefore each one of them should include all relevant
$j$:

\beq E_{\theta\phi}:=\dtp\,|\theta,\phi\9\6\theta,\phi|, \eeq
where
\beq |\theta,\phi\9:=\sum_{j=m}^{N/2}\sqrt{2j+1}\;|j,m(\theta,\phi)\9.
 \label{thphi}\eeq

To verify that this is indeed a POVM, we note that in 
$\int\!E_{\theta\phi}$ there are diagonal terms $(2j+1)|j,m(\theta,\phi)\9
\6j,m(\theta,\phi)|$, which give ${\bf1}_j$, owing to Eq.~(\ref{1j}).
The off-diagonal terms with $j_1\neq j_2$ vanish, as can be seen by
taking their matrix elements between $\6j_1,m_1|$ and $|j_2,m_2\9$
in the standard basis where $J_z$ is diagonal. We have~\cite{edm}

\beq \6j_2,m(\theta,\phi)|j_2,m_2\9=
  {\cal D}^{(j_2)}_{mm_2}(\psi\theta\phi), \eeq
with a similar (complex conjugate) expression for
$\6j_1,m_1|j_1,m(\theta,\phi)\9$. The rotation matrices $\cal D$ are
explicitly given by

\beq {\cal D}^{(j_2)}_{mm_2}(\psi\theta\phi)=
  e^{im\psi}\,d^{(j_2)}_{mm_2}(\theta)\,e^{im_2\phi}, \eeq
where the Euler angle $\psi$ is related to an arbitrary phase which is
implicit in the definition of $|j,m(\theta,\phi)\9$. It is crucial that
a single value of $m$ occurs in all the components of the vectors
$|\theta,\phi\9$ in Eq.~(\ref{thphi}), so that the undefined phases
$e^{\pm im\psi}$ mutually cancel. It then follows from Eq.~(4.6.1) of
Ref.~\cite{edm} that all the off-diagonal matrix elements of
$\int\!E_{\theta\phi}$ vanish, so that we indeed have a POVM.

While Bob's optimal POVM is essentially unique in the
Hilbert space that we have chosen, Alice's signal state, which is

\beq |A\9=\sum_{j=m}^{N/2} c_j\,|j,m(\bn)\9, \eeq
contains unknown coefficients $c_j$. The latter are normalized,

\beq \sum_{j=m}^{N/2} |c_j|^2 = 1, \label{norm} \eeq
but still have to be optimized. 

The probability of detection of the pair of angles $\theta\phi$,
indicated by the POVM component $E_{\theta\phi}$, is

\beq \6A|E_{\theta\phi}|A\9=\dtp\left|\sum_{j=m}^{N/2}c_j\sqrt{2j+1}\,
  \6j,m(\theta,\phi)|j,m(\bn)\9\right|^2. \label{AEA} \eeq
We have~\cite{perel}

\beq \6j,m(\theta,\phi)|j,m(\bn)\9=e^{i\eta}\,d^{(j)}_{mm}(\chi), \eeq
where $\chi$ is the angle between the directions \bn\ and
$\theta\phi$, and the phase $e^{i\eta}$ is related to the arbitrary
phases which are implicit in the definitions of the state vectors in
(\theequation).  The important point is that $e^{i\eta}$ does not depend
on $j$ and therefore is eliminated when we take the absolute value of
the sum in Eq.~(\ref{AEA}). Explicitly, we have

\beq d^{(j)}_{mm}(\chi)=\cos^{2m}(\chi/2)\,P^{(0,2m)}_{j-m}(\cos\chi),\eeq
where $P^{(a,b)}_n(x)$ is a Jacobi polynomial~\cite{perel,edm}. We
shall write $x=\cos\chi$ for brevity, so that the fidelity is

\beq F=(1+\6x\9)/2. \eeq
Our problem is to find the coefficients $c_j$ that maximize $\6x\9$.
Owing to rotational symmetry, we can assume that Alice's
direction \bn\ points toward the $z$-axis, so that $\dtp$ can be
replaced by $dx/2$ after having performed the integration over $\phi$. We
thus obtain

\beq \6x\9=\half\int^1_{-1}xdx\left|\sum_{j=m}^{N/2}c_j\sqrt{2j+1}
 \left({1+x\over2}\right)^mP^{(0,2m)}_{j-m}(x)\right|^2. \eeq

This integral can be evaluated explicitly by using the orthogonality and
recurrence relations for Jacobi polynomials \cite{ortho,recur}. The
result is

\beq \6x\9=\sum_{j,k} c^*_j\,c_k\,A_{jk}, \eeq
where $A_{jk}$ is a real symmetric matrix, whose only nonvanishing
elements are

\beq A_{jj}=m^2/[j(j+1)], \eeq
and
\beq A_{j,j-1}=A_{j-1,j}=(j^2-m^2)/j\sqrt{4j^2-1}. \eeq
The optimal coefficients $c_j$ are the components of the eigenvector
of $A_{jk}$ that corresponds to the largest eigenvalue, and the latter
is $\6x\9$ itself.  The result of the calculation is displayed in Fig.~1
for $m=0$ (which is best) and $m=j$ (which is the method investigated in
Ref.~\cite{mp}). For $m=0$ and large $N$, we find that

\beq 1-F\to5.78317/(N+3)^2. \eeq
This ought to be compared to the result of~\cite{mp}, which was
$1/(N+2)$. For $N=2$ and $m=0$, our result coincides with Ref.~\cite{gp}.
For $N=3$, we obtain $F=0.84495$ with $c_{3/2}=0.60362$ and
$c_{1/2}=0.79755$.  The results for larger $N$ and intermediate values
of $m$ gradually fall between those displayed in Fig.~1. Had we chosen a
definition of fidelity other than Eq.~(1), these results would of course
be different, but the method for solving the problem is in principle the
same.

It thus appears that it is advantageous to take the lowest possible $m$
(namely $m=0$ for even $N$ and $m={1\over2}$ for odd $N$). This is
intuitively quite plausible~\cite{ap}. It would be interesting and
instructive to find a direct proof of Eq.~(\theequation) that does
not rely on a numerical analysis as in the present work.

\bigskip Work by AP was supported by the Gerard Swope Fund and the Fund
for Encouragement of Research. PFS was supported by a grant from the
Technion Graduate School.

\vfill

\noindent FIG. 1. \ $(1-F)$ as a function of $N$. Open circles are for
$m=j$ (Ref.~\cite{mp}), closed circles are for $m=0$ (this work).


\begin{thebibliography}{99}
\bibitem{polyhed}If $N\leq4$, an optimal POVM can have a finite number
of components, corresponding to the vertices of regular polyhedra.
\bibitem{cnvx}A convex combination of POVMs, such as $E_\mu=\sum
w_kE_{k\mu}$, with $w_k>0$ and $\sum w_k=1$, has the physical meaning
that each one of the sets $\{E_{k\mu}\}$ is chosen by the experimenter
with probability $w_k$.
\end{thebibliography}
\end{document}